# Zentropy Theory for Positive and Negative Thermal Expansion

Zi-Kui Liu, Yi Wang, and Shun-Li Shang
Department of Materials Science and Engineering
Pennsylvania State University, University Park, PA 16802, USA

**Abstract**

It has been observed in both natural and man-made materials that volume sometimes decreases with increasing temperature. Though mechanistic understanding has been gained for some individual materials, a general answer to the question "Why does volume sometimes decrease with the increase of temperature?" remains lacking. Based on the thermodynamic relation that the derivative of volume with respect to temperature, i.e., thermal expansion, is equal to the negative derivative of entropy with respect to pressure, we developed a general theory in terms of multiscale entropy to understand and predict the change of volume as a function of temperature, which is termed as zentropy theory in the present work. It is shown that a phase at high temperatures is a statistical representation of the ground-state stable and multiple nonground-state metastable configurations. It is demonstrated that when the volumes of the major nonground-state configurations are smaller than that of the ground-state configuration, the volume of the phase may decrease with the increase of temperature in certain ranges of temperature-pressure combinations, depicting the negative divergency of thermal expansion at the critical point. As examples, positive and negative divergencies of thermal expansion are predicted at the critical points of Ce and $Fe_3Pt$, respectively, along with the temperature and pressure ranges for abnormally positive and negative thermal expansion. The authors believe that the zentropy theory is applicable to predict anomalies of other physical properties of phases because the change of entropy drives the responses of a system to external stimuli.

# 1 Introduction

In solid state physics, thermal expansion is understood by considering the effect of anharmonic terms in the potential energy for a classical oscillator and is thus always positive in the classical region [1]. On the other hand, there are ubiquitous experimental observations of negative thermal expansion (NTE) in both natural materials and human-made materials which have been extensively investigated and reviewed in the literature [2–9]. A range of mechanisms has been developed to explain the NTE characteristics in various materials with their roots tied to lattice vibrational dynamics one way or another, such as the Grüneisen theory of thermal expansion via the tension effect [10,11], particularly at low temperatures. However, those important mechanistic understandings are at the level of interpretation of observations, such as the sign change of Grüneisen parameter [12], and qualitative in nature. Consequently, the fundamental question remains "Why does it happen?", and the lack of fundamental understanding has prevented the development of quantitative predictive approaches.

Nowadays, the lattice vibrational dynamics of a given configuration can be accurately computed by the first-principles calculations based on density functional theory (DFT) [13,14] using advanced computational tools [15–19]. It is even possible to predict the NTE at low temperatures through first-principles phonon calculations of ground-state configurations, such as those of ice and silicon [20] though they cannot be directly applied to phases with their NTE observed at high temperatures. On the other hand, from thermodynamic point of view, the lattice vibrational dynamics is related to entropy [11], and the derivative of volume with respect to temperature, i.e., thermal expansion, is equal to the negative derivative of entropy with respect to pressure [21–23], as follows

$$\left(\frac{\partial V}{\partial T}\right)_P = \frac{\partial^2 G}{\partial T \partial P} = -\left(\frac{\partial S}{\partial P}\right)_T \qquad \textit{Eq. 1}$$

Where $G, V, T, P$, and $S$ are Gibbs energy, volume, temperature, pressure, and entropy of the phase, respectively. It is also known thermodynamically that thermal expansion along with all other physical properties in terms of the second derivatives of free energy diverges at a critical point, i.e., the limit of stability of a macroscopically homogeneous phase [12,21,23]. The key challenge is thus to predict the entropy of a phase as a function of temperature and pressure and its divergency at a critical point.

Recently, we discussed the fundamentals of thermal expansion and thermal contraction including a predictive approach in terms of DFT-based first-principles calculations [24] and demonstrated its applicability in predicting both positive and negative infinite thermal expansions at the critical points of $Ce$ and $Fe_3Pt$, respectively, with inputs solely from first-principles calculations and without fitting parameters [20], but did not clearly address the above question. The present work aims to examine the volume and thermal expansion from the thermodynamic perspectives as shown by Eq. 1 in terms of intrinsic multiscale characteristics of entropy based on statistical mechanics [23,25]. The authors hope that the combination of mechanistic and thermodynamic perspectives [26] can present a general theory for a more comprehensive answer to the question presented above. In the present paper, "configurations" and "states" are used interchangeably to denote possible stable or nonstable configurations of a system with all its internal variables specified.

## 2  Fundamentals of entropy and the zentropy theory

Entropy is a thermodynamic quantity representing the possible configurations in a system, and the second law of thermodynamics stipulates that when a system is disturbed from its equilibrium state, one or more irreversible internal processes will take place in the system and must result in an increase of entropy, i.e., the positive entropy production [21,22]. The volume change of a system with respect to the injection of heat from its surroundings is an irreversible internal process and must thus result in a positive entropy production, which will be further discussed in a separate paper on cross phenomena involving multiple internal processes.

Discussion of entropy usually starts with the configurational entropy at a specific scale of interest as follows

$$S^{conf} = -k_B \sum_k p_k ln p_k \qquad Eq.\ 2$$

where $p_k$ is the probability of configuration $k$ at the scale of the observation, and $k_B$ the Boltzmann constant. It is evident that the total entropy of the system needs to include the entropy of each configuration and the configurational entropy among them as follows [25]

$$S = \sum_k p_k S_k + S^{conf} = \sum_k p_k (S_k - k_B ln p_k) \qquad Eq.\ 3$$

where $S_k$ is the entropy of configuration $k$ which can be further decomposed into configurations in the lower scales with the same formula as Eq. 3. The scale higher than the observation is usually considered as the surroundings of the system [21,22], which dictates the statistical ensemble to be used to study the system with some typical ones as follows

(i) The microcanonical ensemble under constant mass ($N$), volume ($V$), and the total energy in the system ($E$) without any exchanges between the system and the surrounding ($NVE$ ensemble, with entropy as the characteristic state function [21]);

(ii) The canonical ensemble under constant mass ($N$), volume ($V$), and temperature ($T$) with only heat exchange between the system and the surrounding ($NVT$ ensemble, with Helmholtz energy, $F$, as the characteristic state function [21]);

(iii) The grand canonical ensemble under constant chemical potential ($\mu$), volume ($V$), and temperature ($T$) with both mass and heat exchanges between the system and the surrounding ($\mu VT$ ensemble, with $\Theta = F - \sum \mu_i N_i$ as the characteristic state function [21]);

(iv) The isothermal–isobaric ensemble under constant mass ($N$), pressure ($P$), and temperature ($T$) with both volume and heat exchanges between the system and the surrounding ($NPT$ ensemble, with Gibbs energy, $G$, as the characteristic state function [21]);

(v) The isoenthalpic-isobaric ensemble under constant mass ($N$), pressure ($P$), and enthalpy ($H$) with exchange of volume but without heat exchange between the system and the surrounding ($NPH$ ensemble, also with entropy as the characteristic state function [21]); and

(vi) The partial grand isothermal–isobaric ensemble under constant chemical potentials for some components ($\mu_i$), constant mass for other components ($N_j$), and constant pressure ($P$) and temperature ($T$) ($\mu_i N_j PT$ ensemble, with $\Phi = G - \sum_{i<j} \mu_i N_i$ as the characteristic state function).

Ensemble (vi) is useful when mass exchanges between the system and the surroundings occur only for some, but not all components such as systems under consideration of carburization or oxidation. As common experiments are conducted under the *NPT* ensemble, i.e., ensemble (iv) above, Gibbs energy, is the characteristic state function to describe the property of the system and can be written as follows

$$G = E - TS + PV = \sum_k p_k E_k - T \sum_k p_k (S_k - k_B \ln p_k) + P \sum_k p_k V_k$$
$$= \sum_k p_k G_k + k_B T \sum_k p_k \ln p_k$$

*Eq. 4*

where $E$, $V$, $G_k$, $E_k$, and $V_k$ are the internal energy, and volume of the system and the configuration $k$, respectively. It can be seen that the Gibbs energy of the system consists of weighted linear combination of the Gibbs energies of its individual configurations and their nonlinear statistical interactions through the configurational entropy at the observation scale. It is noted that the last term in Eq. 4 is usually missing in the literature, which is the key feature of the present theory.

From the definition of partition function, the following equation can be obtained

$$Z = e^{-\frac{G}{k_B T}} = \sum_k e^{-\frac{G_k}{k_B T}} = \sum_k Z_k \qquad \textit{Eq. 5}$$

where $Z$ and $Z_k$ are the partition functions of the system and the configuration $k$ in the *NPT* ensemble, respectively. It is noted that when our approach was originally developed [20,27–29], we started with the postulation of Eq. 5 and used the Helmholtz energy by considering the *NVT* ensemble. While in the present work as shown above, Eq. 5 can be rigorously derived from Eq. 3 as originally presented in Ref. [25].

The partition function for the *NVT* ensemble commonly uses $E_k$ instead of Helmholtz energy of each configuration ($F_k$) in our previous works [20,27–29]. This reflects the important difference in terms of entropy, i.e., when $E_k$ is used, the entropy is represented by Eq. 2; while when $F_k$ or $G_k$ is used, the entropy is represented by Eq. 3. As mentioned above, the entropy by Eq. 2 is for configurations at one scale only, while the entropy by Eq. 3 is for the total entropy of the system, containing contributions from this scale and all lower scales, i.e., multiscale. Since our multiscale formulism is closely related to the partition function denoted by $Z$ in Eq. 5, we were suggested to term our approach as the zentropy theory during a seminar [30] with $z$ representing the partition function (i.e., the zustandssumme in German coined by Max Planck) with its meaning of "sum over states".

One significance of the zentropy theory is the expression for the probability of each configuration at the scale of observation as follows

$$p_k = \frac{Z_k}{Z} = e^{-\frac{G_k - G}{k_B T}} \qquad Eq.\ 6$$

which shows that the probability of a configuration is related to the free energy difference between those of the configuration and the system. While conventionally, the difference would be between the internal energy of the configuration and the free energy of the system, which omits the entropy of individual configurations and is thus less accurate.

The entropy of a given configuration can be routinely predicted by DFT-based first-principles calculations [18], which can be performed using the recently developed high throughput DFT Tool Kit (DFTTK) [19,31], as follows

$$S_k = S_{k,el} + S_{k,vib} \qquad Eq.\ 7$$

where $S_{k,el}$ and $S_{k,vib}$ are the entropies of configuration $k$ due to thermal electrons and lattice vibrations or phonons, respectively. The Gibbs energy of configuration $k$ can then be obtained as

$$G_k = E_{k,c} + F_{k,el} + F_{k,vib} + PV_k \qquad Eq.\ 8$$

where $E_{k,c}$ is the 0 K static total energy, $F_{k,el}$ the thermal electronic contribution, and $F_{k,vib}$ the vibrational contribution, all as a function of $V_k$ with $F_{k,el}$ and $F_{k,vib}$ also as a function of $T$.

## 3  Volume of a system

The volume of a system can be obtained as follows

$$V = \frac{\partial G}{\partial P} = -k_B T \frac{\partial lnZ}{\partial P} = -\frac{k_B T}{Z} \sum_k \frac{\partial Z_k}{\partial P} = \sum_k p_k V_k \qquad Eq.\ 9$$

which was already used in Eq. 4. With the ground-state stable configuration denoted by $g$, Eq. 9 can be re-organized as

$$V = V_g + \sum_k p_k (V_k - V_g) \qquad Eq.\ 10$$

One can immediately observe from Eq. 10 that the negative values of $V_k - V_g$ could result in the decrease of the system volume with the increase of temperature if the decrease due to $\sum p_k (V_k - V_g)$ is more than the increase of $V_g$ with respect to temperature. It is thus self-evident that the necessary condition for the volume of the system to decrease with temperature is that the major nonground configurations with $V_k < V_g$ have relatively high probabilities with increasing temperature, as determined by Eq. 6.

While the entropy of the ground-state stable configuration can be accurately predicted by the quasiharmonic approximations (QHA) through DFT-based first-principles calculations (see Eq. 8) [18], the anharmonicity of the system is unavoidable at high temperatures due to the interference of other nonground-state metastable configurations through the nonlinear configurational entropy among them (see Eq. 3) as their probability is significantly increased with respect to temperature (see Eq. 6). From thermodynamics [21–23], it is known that the limit of anharmonicity is at a critical point where the system reaches its limit of stability with all its properties diverged[21–23], i.e.,

$$\frac{\partial S}{\partial T} = \frac{\partial V}{\partial (-P)} = +\infty \qquad Eq.\ 11$$

The positive sign is because $S$ and $T$ are conjugate variables in the combined law of thermodynamics, and so are $V$ and $-P$ [21–23].

It is evident that the derivative of volume with respect to temperature, i.e. $\frac{\partial V}{\partial T}$, also diverges at the critical point, but the thermodynamic stability criterion does not require $\frac{\partial V}{\partial T}$ to be positive [21–23]. From Eq. 10 and discussion above, one can thus conclude the following at the critical point

$$\frac{\partial V}{\partial T} = +\infty \ \ \text{when}\ V_k > V_g$$
$$\frac{\partial V}{\partial T} = -\infty \ \ \text{when}\ V_k < V_g \qquad Eq.\ 12$$

The answer to the question "Why does volume sometimes decrease with the increase of temperature?" is thus that the nonground-state metastable configurations with increased statistical probability at high temperatures have their volumes smaller than that of the ground-state stable configuration, and the volume of the system is the weighted sum of volumes of individual configurations as shown by Eq. 10.

## 4   Examples: $Ce$ and $Fe_3Pt$

In our previous works, the temperature and pressure phase diagrams of $Ce$ [27,28] and $Fe_3Pt$ [29] were predicted by the zentropy theory including their critical points. In $Ce$, three configurations were considered, i.e., the nonmagnetic (NM), antiferromagnetic (AFM), and ferromagnetic (FM) [28]; while in $Fe_3Pt$, $2^9 = 512$ magnetic spin configurations were considered with nine Fe atoms in the supercell for DFT-based calculations, resulting in 37 unique spin-flip configurations (SFC) [29]. Their 0 K static energies, i.e., the $E_{k,c}$ in $Eq.$ 8, are plotted in Figure 1a and b, respectively. It can be seen that in $Ce$, the volume of the ground-state stable NM configuration ($\alpha$-Ce) is smaller than those of the nonground-state metastable AFM and FM ($\gamma$-Ce) configurations, while in $Fe_3Pt$, the volume of the ground-state stable FM configuration is larger than all other nonground-state metastable SFCs. Their Helmholtz energies as a function of temperature under ambient pressure, equal to their Gibbs energies, are shown in Figure 2.

The predicted temperature-pressure phase diagrams of $Ce$ [28] and $Fe_3Pt$ [29] are shown in Figure 3. The lines represent the conventionally defined two-phase equilibrium regions which are one-dimensional based on the Gibbs phase rule [21,22], i.e., the low temperature NM and high temperature FM phases for Ce, and low temperature FM and high temperature paramagnetic (PM) phases for $Fe_3Pt$, respectively, though each of them is a statistical mixture of all configurations with different statistical probabilities. The probabilities of various configurations as a function of temperature are plotted in Figure 4 for $Ce$ and $Fe_3Pt$, respectively. The second-order transition temperature is defined when the probability of the ground-state stable configuration equals to the sum of nonground-state metastable configurations; similar results are obtained if the maximum heat capacities are used. The slops of the two-phase equilibrium region shown in Figure 3 are positive for $Ce$ and negative for $Fe_3Pt$ and are related to the volume and entropy differences of the ground-state stable configuration and nonground-state metastable configurations in terms of the Clausius-Clapeyron equation as follows

$$\frac{\partial T}{\partial P} = \frac{\Delta V}{\Delta S} \qquad Eq.\ 13$$

where $\Delta V$ and $\Delta S$ are the volume and entropy differences between the ground-state stable configuration and nonground-state metastable configurations. Since the entropy increases with temperature, the sign of Eq. 13 for the slope of the two-phase region is determined by the volume difference between the ground-state stable configuration and nonground-state metastable configurations in accordance with Eq. 12 as follows

$$\frac{\partial T}{\partial P} > 0 \ \ \text{when}\ V_k > V_g$$
$$\frac{\partial T}{\partial P} < 0 \ \ \text{when}\ V_k < V_g \qquad Eq.\ 14$$

Therefore, the negative slope of the two-phase equilibrium line in the temperature-pressure potential phase diagram provides a useful indication for volume to decrease with temperature in the system. This criterion was used to predict the potency of negative thermal expansion based on available temperature-pressure phase diagrams with remarkable agreement with available experimental observations [32]. Since there are many two-phase equilibrium lines with negative slope in temperature-pressure phase diagrams, the negative thermal expansion phenomenon is much more common than one typically thinks.

This is further demonstrated by replacing the pressure in the temperature-pressure potential phase diagram by its conjugate molar quantity, volume, which resulted in the temperature-volume mixed potential-molar quantity phase diagrams shown in Figure 5 for $Ce$ and $Fe_3Pt$, respectively. The two-phase equilibrium line in the temperature-pressure potential phase diagram now becomes a two-dimensional area as a miscibility gap between two phases, i.e., from one degree of freedom (number of independent potentials) in a potential phase diagram to two dimensions in a phase diagram (number of independent variables) with one potential and one molar quantity. As Hillert discussed in detail [21], Gibbs phase rule concerns the number of potentials that can change independently without changing the number of phases in equilibrium and needs to be modified when it is applied to phase diagrams with molar quantities as axis variables. There two phases are dominated by the ground-state stable configuration (low temperature phase) and nonground-state metastable configurations (high temperature phase), respectively. These two phases merge into a single phase at the critical point. Figure 5 also includes curves of isobaric volume as a function of temperature. The divergency in accordance with Eq. 12 at the critical point is clearly shown. Furthermore, the decrease of volume with respect to temperature in $Fe_3Pt$ under various pressures is marked by the purple diamond symbols. While for $Ce$, the purple diamond symbols denote the abnormally large increase of volume with respect to temperature.

As mentioned in the introduction, at low temperatures, the nonground-state metastable configurations can be accessed by phonon vibrations of the ground-state stable configuration. Consequently, the phonon calculations of the ground-state stable configuration will include the contributions from the nonground-state metastable configurations and can thus predict the decrease of volume with respect to temperature at temperature close to 0 K by phonon calculations alone, which were demonstrated for ice (H₂O) and Si as shown in Figure 6 [20].

## 5   Summary

The zentropy theory is discussed in the present work in the framework of Boltzmann-Gibbs entropy formalism by considering multiscale entropic contributions to a phase at finite temperatures in terms of thermal electronic, vibrational, and spin configurations. It demonstrates that a phase at finite temperatures is composed of multiple configurations including both the ground-state stable configuration and nonground-state metastable configurations. The answer to the question "Why does volume sometimes decrease with the increase of temperature?" is that when the volumes of major nonground-state metastable configurations are smaller than that of the ground-state stable configuration, the volume of the phase may decrease with the increase of temperature in a range of temperature and pressure combination. This occurs when the decrease of volume due to the replacement of the ground-state stable configuration by the nonground-state metastable configurations is more than the increase of volume of the ground-state stable configuration. The change of volume diverges at the limit of stability of the phase based on thermodynamics is confirmed by the predictions of the zentropy theory for both $Ce$ and $Fe_3Pt$ at their critical points where the divergence is positive for Ce and negative for Fe₃Pt, respectively. The present zentropy theory provides a thermodynamic framework for mechanistic understanding of thermal expansion and has the potential to predict anomalies of other physical properties such as bulk modulus, heat capacity, and the order-disorder transition temperature in terms of the second derivatives of free energy for discovering materials with emergent behaviors. We are also actively working on using the zentropy theory to predict ferroelectric and superconducting transitions.

## 6  Acknowledgements

The work presented in this paper came from many projects supported by funding agencies in the United States in last two decades with the latest ones including the National Science Foundation (NSF, with the latest Grants CMMI-1825538 and CMMI-2050069), Department of Energy (with the latest Grants being DE-FE0031553, DE-NE0008757, DE-EE0008456, DE-SC0020147, and DE-AR0001435), NASA Space Technology Research Fellowship (with the latest Grant 80NSSC18K1168 ), Army Research Lab (with the latest Grant W911NF-14-2-0084), Office of Naval Research (with the latest Grant N00014-17-1-2567), Wright Patterson AirForce Base, NASA Jet Propulsion Laboratory, and the National Institute of Standards and Technology, plus a number of national laboratories and companies that supported the NSF Center for Computational Materials Design (NSF, 0433033, 0541674/8, 1034965/8), the Roar supercomputer at the Pennsylvania State University, the resources of NERSC supported by the Office of Science of the U.S. Department of Energy under contract No. DE-AC02-05CH11231, and the resources of XSEDE supported by NSF with Grant ACI-1053575.  The authors would like to thank numerous collaborators over the years as reflected in the publications listed in ref. 14. ZKL would like to thank Josiah Roberts from University of Buffalo for suggesting the term zentropy during a seminar. The preparation of manuscript is partially supported by the Endowed Dorothy Pate Enright Professorship at The Pennsylvania State University.  The proper copyrights for all the figures are granted by respective publishers.

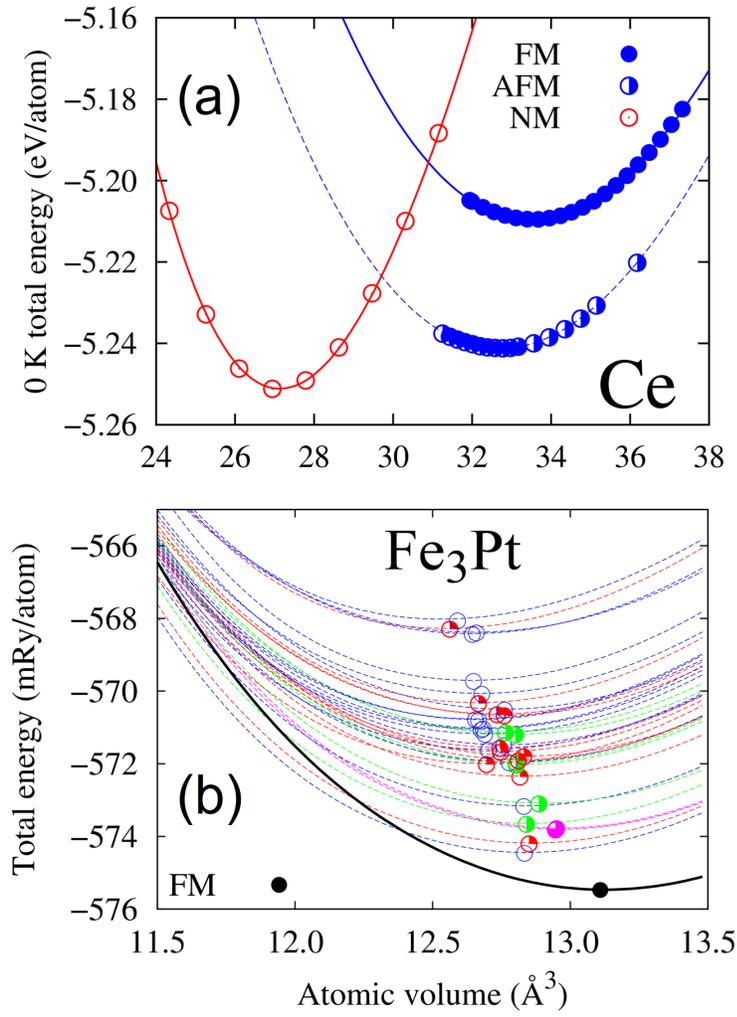

Figure 1: 0 K static energies of (a) Ce [28] and (b) Fe$_3$Pt [29] from DFT-based first-principles calculations. The symbols in figure (a) are the DFT-based predictions and in (b) correspond to equilibrium energy and equilibrium volume for each configuration.

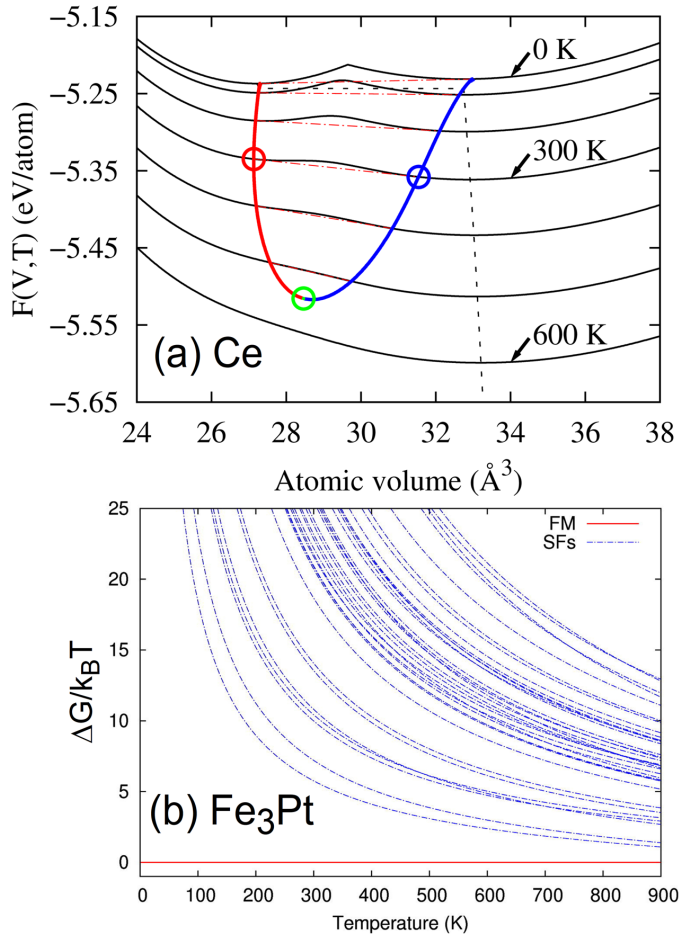

*Figure 2: Helmholtz energy as a function of temperature under ambient pressure for (a) Ce [28] and (b) Relative Gibbs energy (ΔG/k$_B$T) under ambient pressure for Fe$_3$Pt [32] from DFT-based first-principles calculations. In figure (a), the red and blue symbols at 300 K are on the common tangent to determine phase boundary, while the green symbol corresponds to the critical point.*

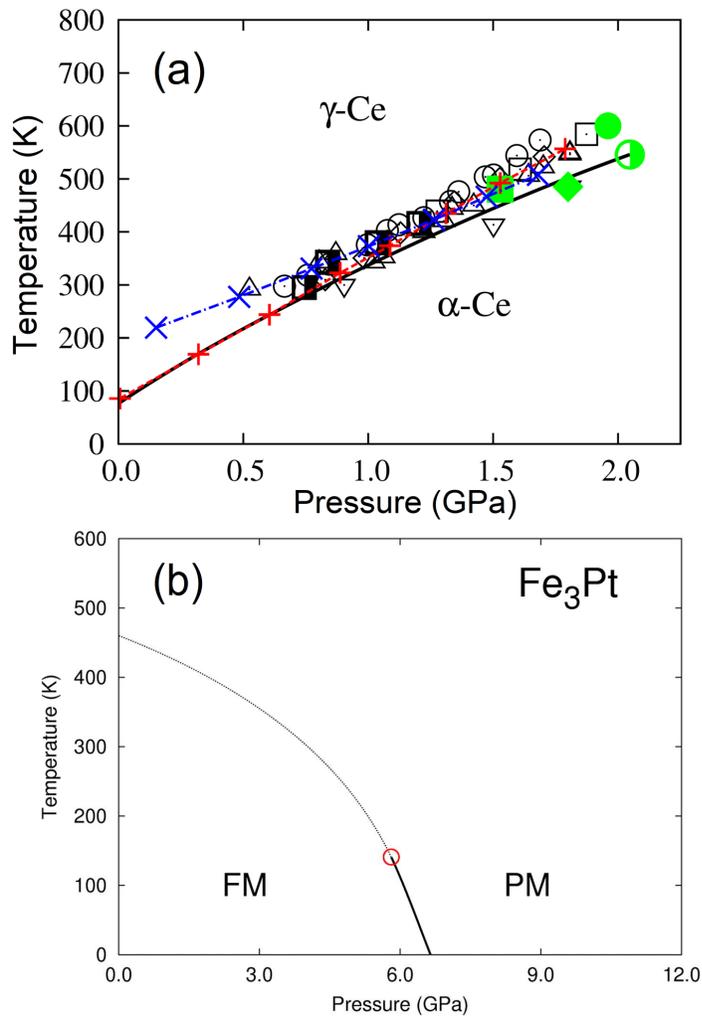

*Figure 3: Predicted temperature-pressure phase diagrams of (a) Ce [28] and (b) $Fe_3Pt$ [29] in terms of the* zentropy *approach. In figure (a) the symbols are experimental data and the red circle in (b) corresponds to the critical point.*

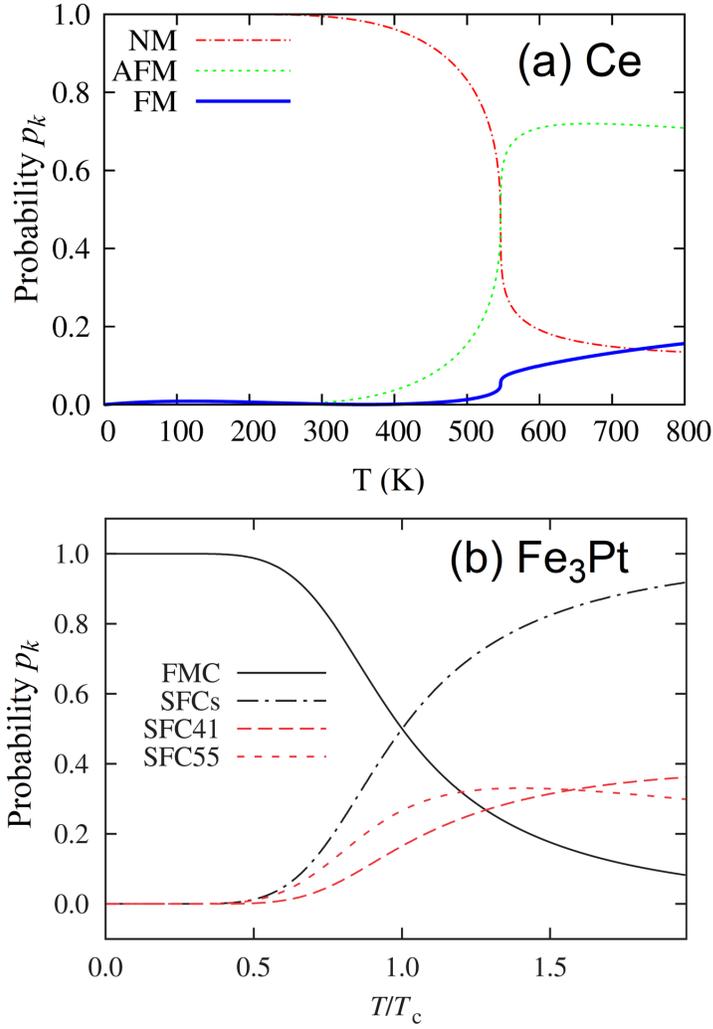

*Figure 4: Predicted probabilities of various configurations as a function of temperature for (a) Ce [28] at 2.05 GPa near the critical point and (b) $Fe_3Pt$ [29] at ambient pressure far away from its critical point. Here, NM represents nonmagnetic configuration, AFM antiferromagnetic configuration, FM (or FMC) ferromagnetic configuration, and SFC(s) spin-slip configuration(s).*

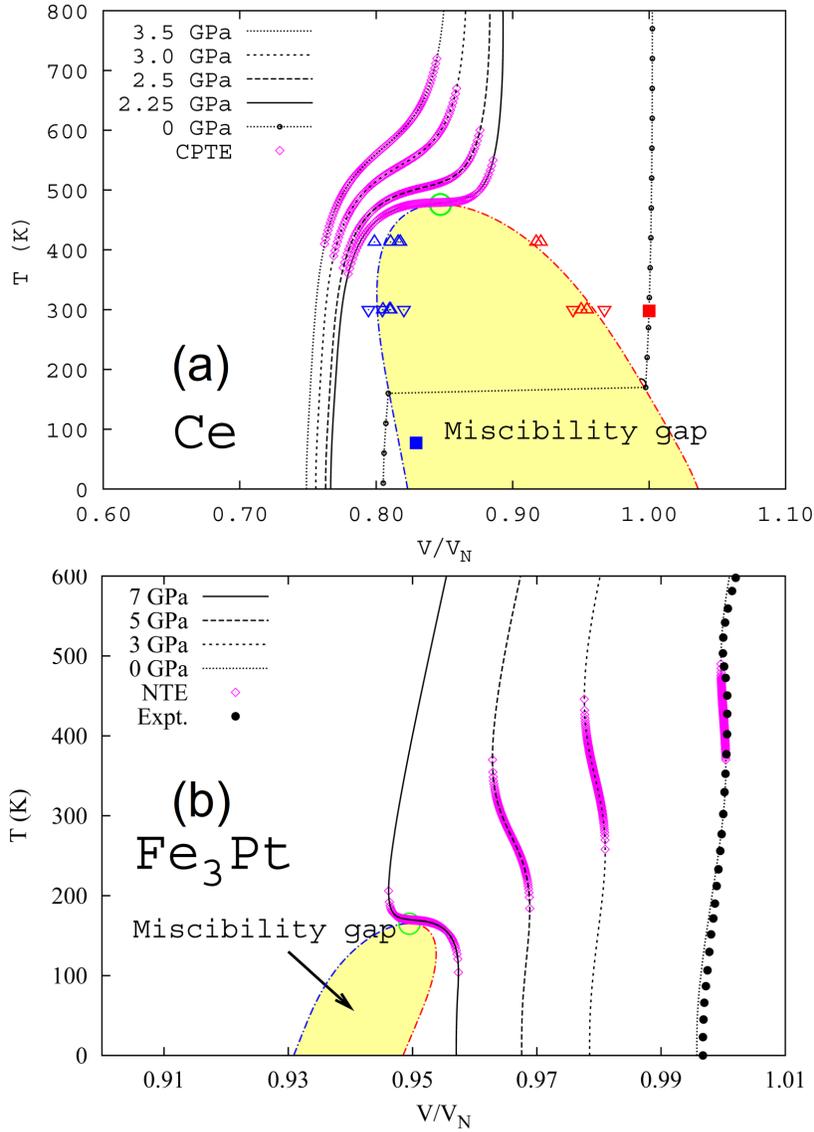

*Figure 5: Predicted T-V phase diagrams of (a) Ce and (b) Fe$_3$Pt with isobaric volume curves with the volume, V, normalized to V$_N$ at 298 K and 1 atm. The purple diamonds mark the anomalous regions of colossal positive or negative thermal expansions (CP/NTEs) in Ce and Fe$_3$Pt, respectively, including the divergences at the critical point by green circle [20]. The other symbols are experimental data points.*

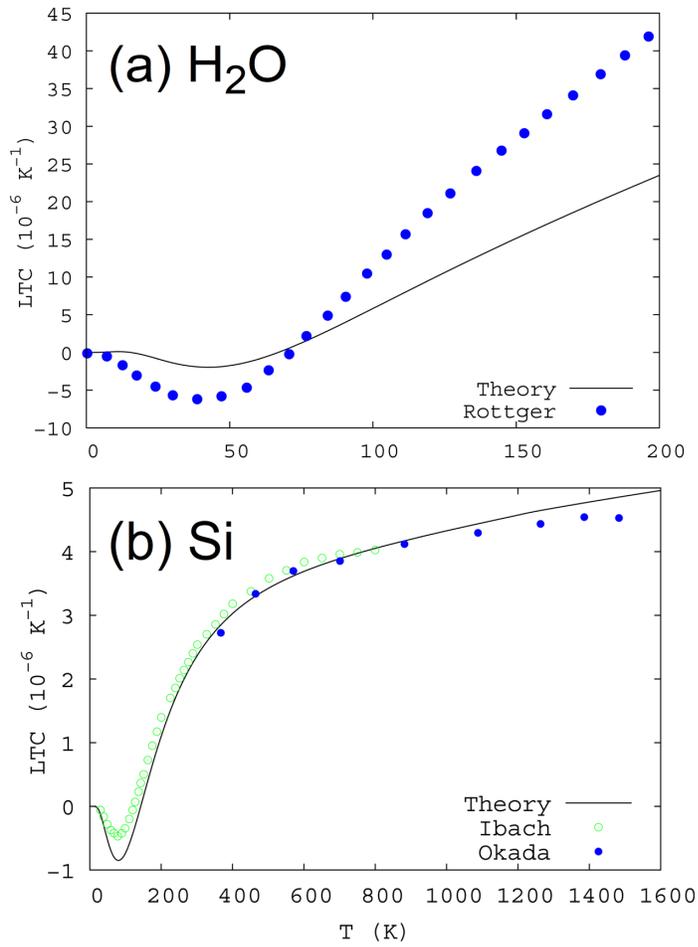

*Figure 6: Linear thermal expansions of (a) Ice ($H_2O$) and (b) Si from DFT-based phonon calculations in combination with experimental data (symbols)* [20].